\documentclass[aps,twocolumn,pra,showpacs]{revtex4}
\usepackage[dvips]{graphicx}
\usepackage{dcolumn}
\usepackage{bm}

\begin{document}

\author{Jakub S. Prauzner-Bechcicki$^1$, Krzysztof Sacha$^1$, Bruno Eckhardt$^2$, and Jakub Zakrzewski$^1$}
\affiliation{$^1$Instytut Fizyki imienia Mariana Smoluchowskiego,
  Uniwersytet Jagiello\'nski,
 ulica Reymonta 4, PL-30-059 Krak\'ow, Poland \\
$^2$Fachbereich Physik, Philipps-Universit\" at Marburg, D-35032
Marburg, Germany}

\title{Non-sequential double ionization of molecules}

\date{\today}

\begin{abstract}
Double ionization of diatomic molecules by short linearly polarized laser pulses is analyzed.
We consider the final stage of the ionization process, that is the decay of a highly excited
two electron molecule, which is formed after re-scattering.
The saddles of the effective adiabatic potential energy close to which
simultaneous escape of electrons takes  place are identified. Numerical simulations of the ionization  of molecules
show that the process can be dominated by 
either sequential or non-sequential events. 
 In order to increase the ratio of non-sequential to
sequential ionizations very short laser pulses should be applied.
\end{abstract}

\maketitle

\section{Introduction}
High intensity ultrashort-pulse lasers allow experimental studies of multi-electron
effects, such as a non-sequential double ionization, high order harmonic generation or
above threshold ionization \cite{silap93,silap00}. While the single ionization of atoms or molecules, as
well as the high order harmonic generation can be described within 
a single active
electron model, such an approximation in the case of double ionization and
laser intensity below the saturation value gives
ionization rates that are much smaller than experimentally observed,
indicating that interactions between electrons are important
\cite{silap93,silap00,schafer93,yang93}. 
In this paper we
consider the double ionization of molecules within a classical model for electrons
in a combined Coulomb and external field applying the approach developed
in~\cite{eckhardt01pra1,eckhardt01pra2,eckhardt01epl,eckhardt03jpb} for the multiple ionization of atoms.

Multi-photon double ionization of atoms in strong laser fields can be regarded as a three step
process \cite{silap93,silap00}. In the first step one electron tunnels out through the Stark saddle and then is
returned back to the nucleus \cite{corkum93,kulander93}. Thereafter, in the second step a highly excited state of an atom or a
molecule is formed at the expense of the energy brought back by the returning electron 
(up to $3.17U_p$, where $U_p$ is a ponderomotive energy). Finally, in the third step such a highly
excited compound state decays in several ways through a single, double or multiple ionization.
The starting point in our classical analysis of the double ionization of molecules
is the excited complex, i.e. we assume that we have an initial state of two
highly excited electrons close to the molecular core. 
We shall focus on double ionization events that can then appear as a possible 
channels of decay. Among those events there are: a sequential process
(when the highly excited compound state decays emitting electrons one by one) 
and a non-sequential event (two electron leave the excited molecule simultaneously).
   
There have been experimental research aimed on double ionization of diatomic
molecules~\cite{cornaggia98,guo98,guo00,eremina04} which showed that there are
differences between molecular species. For example, nitrogen molecule, 
$\rm N_2$, 
clearly exhibits a ''knee-structure'' in a double-to-single
ionization yields ratio, like atoms do, while for oxygen molecule, $\rm O_2$, 
such a structure is rather not visible \cite{guo98,guo00}. Moreover, 
in the case of $\rm N_2$ it seems, that electrons escape
with similar momenta along field polarization axis more often than in the case
of $\rm O_2$ \cite{eremina04}. We consider this problem from the point 
of view of the classical analysis, which starts after the formation of the highly
excited compound state, and discuss possible reasons for such different
experimental observations. 

Paper is structured as follows in Sec.~\ref{model} we introduce a theoretical model on which we based our
analysis. Sec.~\ref{local} consists of the identification and description of the saddles in the
potential. Then in Sec.~\ref{sym} we present results of the numerical simulations. And finally, in
Sec.~\ref{summary} we conclude.

\section{Model}
\label{model}
As we mentioned above, in our starting situation 
(as a result of the former re-scattering)
there are two highly excited electrons 
close to the molecular core 
in the presence of linearly polarized laser field. The motion of the molecular core is
frozen due to the fact that for short laser pulses molecules have not 
enough time to change their orientation \cite{eremina04,miyazaki04}. 
Thus, the Hamiltonian reads (in atomic units):
 \begin{equation}
 H=\frac{{\bf p}_1^2+{\bf p}_2^2}{2}+V,
 \end{equation}
 where the potential,
 \begin{equation}
 V=V_1+V_2+V_{12}+V_F,
 \end{equation}
 consists of the potential energies associated with interactions of the electrons 
 with the nuclei (the entire structure 
 of the molecular core is approximated 
 by two positively charge nuclei),
 \begin{equation}
 V_i=-\frac{1}{|{\bf r}_i-{\bf R_1}|}-\frac{1}{|{\bf r}_i-{\bf R_2}|},
 \end{equation}
 for ($i=1,2$), where ${\bf R}_i$ and ${\bf r}_i$ indicate position 
 of the nuclei and the electrons,
 respectively, the repulsion between electrons,
 \begin{equation}
 \label{pot2}
 V_{12}=\frac{1}{|{\bf r}_1-{\bf r}_2|},
 \end{equation}
 and the term describing interaction with the field (polarized along the $z$ axis),
 \begin{equation}
 V_F=F(t)(z_1+z_2).
 \end{equation}
 The electric field strength $F(t)$ has an oscillatory component times 
 the envelope from the pulse,
 namely:
 \begin{equation}
 F(t)=Ff(t)\cos(\omega t + \phi),
 \end{equation} 
with $F$, $\omega$ and $\phi$ the peak amplitude, frequency, and initial phase of the filed,
respectively, and with 
\begin{equation}
f(t)=\sin^2(\pi t/T_d),
\end{equation}
the pulse envelope of duration 
\begin{equation}
T_d=n\frac{2\pi}{\omega}, 
\end{equation}
where $n$ is number of cycles in the pulse.

For a sake of convenience, we shall place the origin of our coordinate system in the center of mass of
the nuclei, then two parameters appear, namely: $d$, distance between nuclei, and $\theta$, angle between
molecular axis and $z$ axis (polarization axis). Without loss of generality we assumed that the molecule lies in $xz$ plane, hence
potential energies for each electron is:
\begin{eqnarray}
\label{pot1}
V_i=-\frac{1}{\sqrt{(x_i+\frac{d}{2}\sin \theta)^2+y_i^2+(z_i+\frac{d}{2}\cos \theta)^2}}\cr
-\frac{1}{\sqrt{(x_i-\frac{d}{2}\sin \theta)^2+y_i^2+(z_i-\frac{d}{2}\cos \theta)^2}}.
\end{eqnarray}

As one can see, in our model the only difference between different 
diatomic species lies in the distance, $d$, between the nuclei. 

\section{Local analysis}
\label{local}
What we consider in this paper is the evolution of electrons in combined 
Coulomb and external fields after a re-scattering event when the highly excited compound state is formed.
For one electron in the Coulomb potential, if the external field is non-zero the Stark saddle is
opened. Then the electron can get away through this saddle and ionize.
In our case we have two electrons and if there were no interaction between the electrons they
could escape simultaneously through the same saddle on
top of each other. But once there is repulsion between electrons, the Stark saddle splits into two saddles that lie on the opposite sides of the field polarization axis. In the case of atoms,
 those two saddles lie
symmetrically with respect to the polarization axis. Then the motion of the electrons can be confined in
some symmetry subspace~\cite{eckhardt01pra1,eckhardt01pra2}. 
For diatomic molecules (such as $\rm N_2$ or $\rm O_2$)
that does not generally occur since they possess their own symmetry axis which can be oriented at any angle to the
polarization axis
destroying the global symmetry. Nevertheless, saddles are formed by
the external field and following arguments presented
in~\cite{eckhardt01pra1,eckhardt01pra2,eckhardt01epl,eckhardt03jpb} we assume that electrons  to leave the
molecule in a correlated manner have to pass simultaneously close to them.

Before the double ionization escape both electrons  pass close to the nuclei  where they
interact strongly with each other and with the nuclei. For that reason we assume that all memory of the previous motion is lost. Then it is
correct to assume  that the compound state which decays to a doubly charged molecule may be classically simulated by 
a statistical distribution for  two electrons close to the nuclei. Furthermore, the classical motion of
the electrons 
is fast compared to the field oscillations and an adiabatic approximation, keeping  the field fixed, becomes 
useful in the analysis of the ionization channels. 
Using this adiabatic assumption we will identify and describe saddle points, through which the electrons can escape. 
  
\begin{figure}[ht]
\includegraphics[height=0.3\textwidth,width=0.4\textwidth]{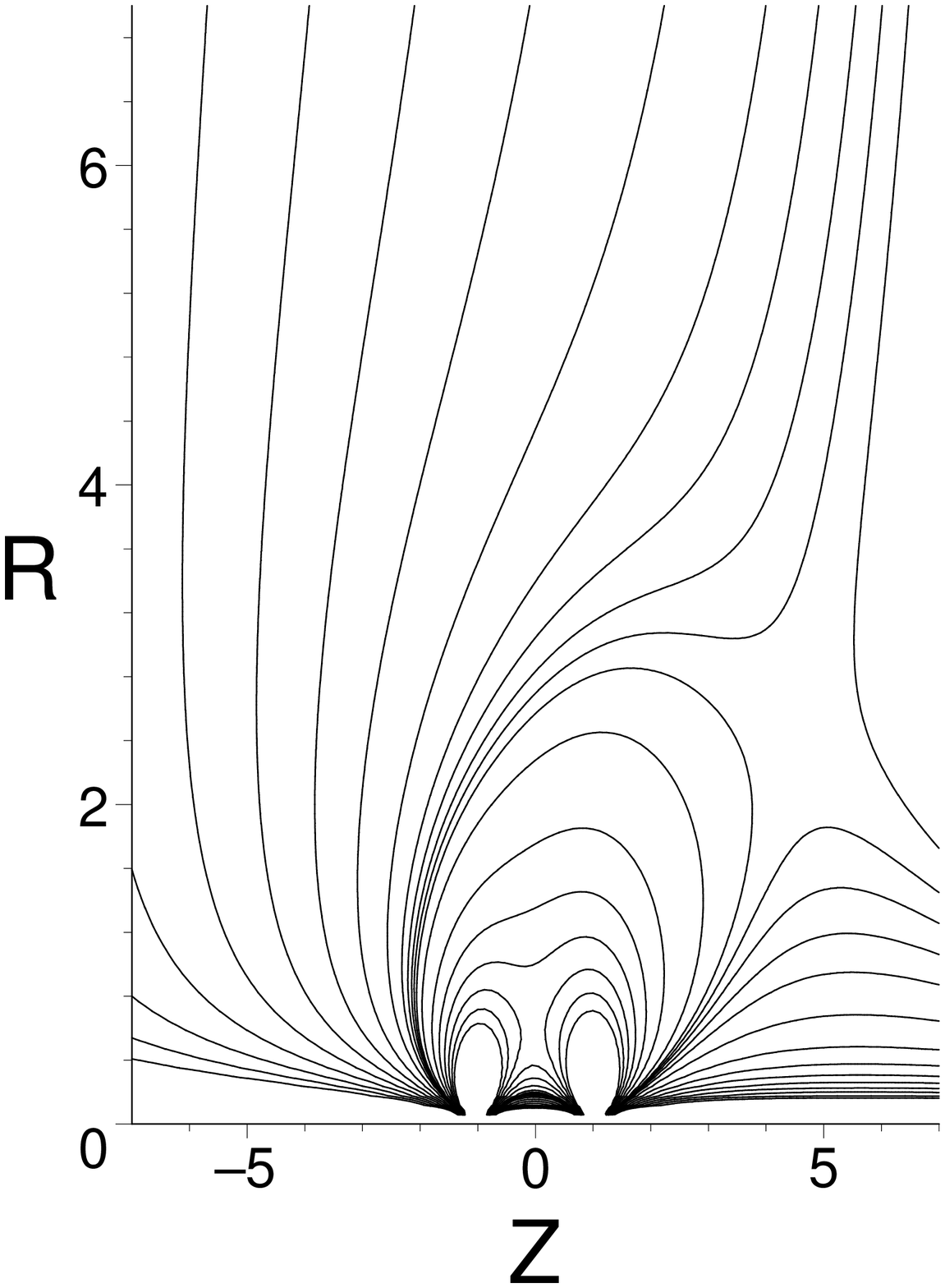}
\includegraphics[height=0.3\textwidth,width=0.4\textwidth]{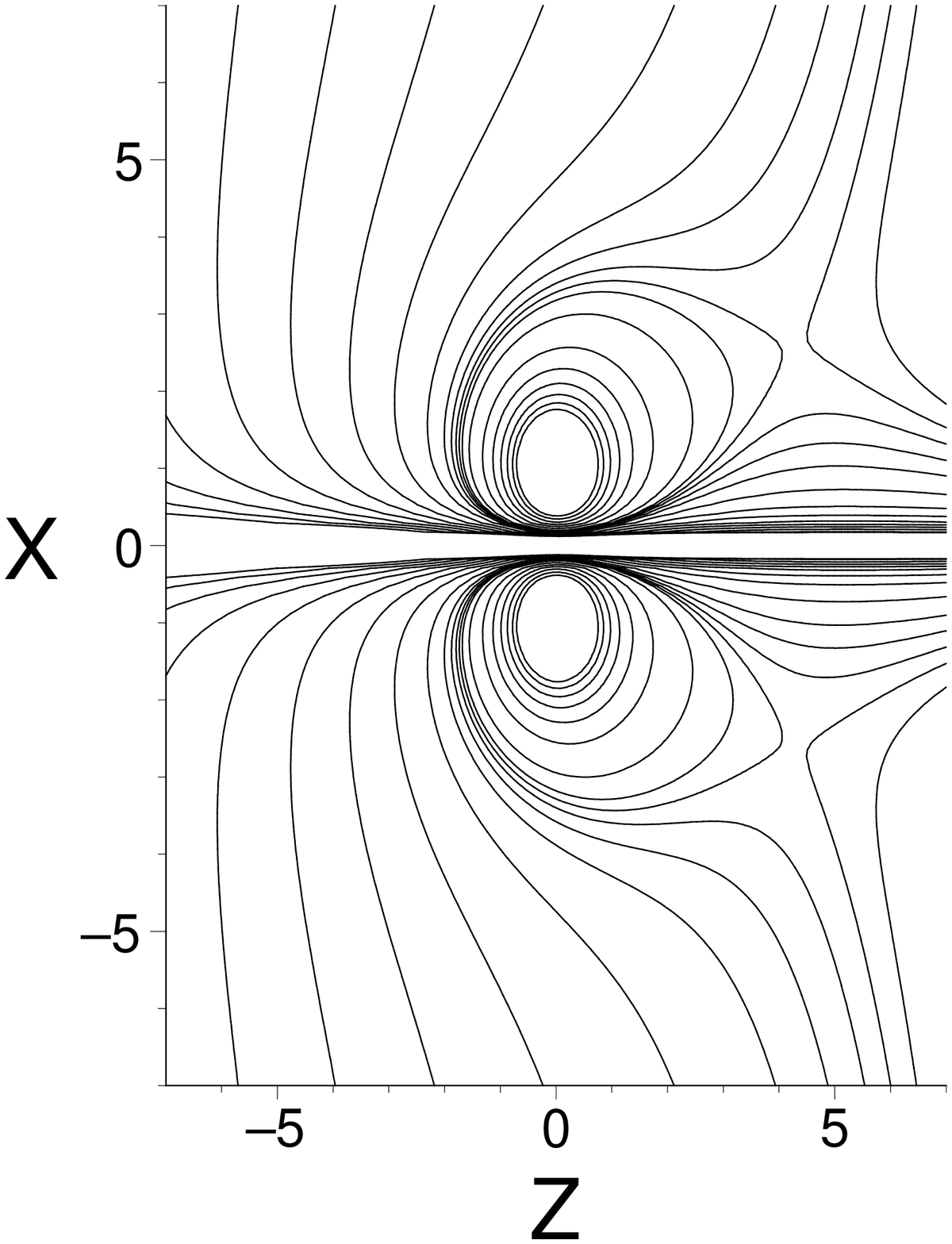}
\includegraphics[height=0.3\textwidth,width=0.4\textwidth]{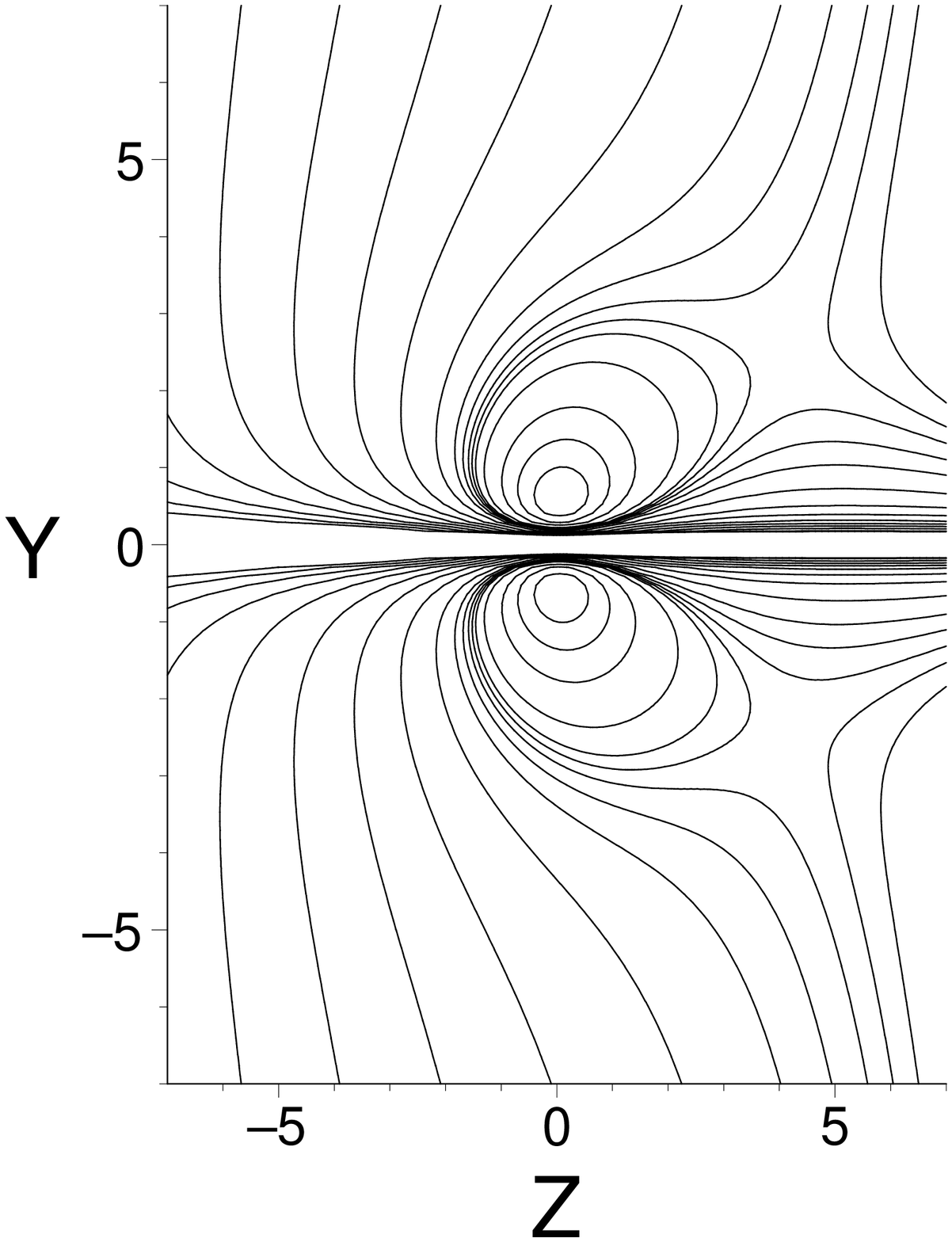}
\caption{Section through equipotent surfaces of the adiabatic potential,
$V=V_1+V_2+V_{12}+V_F$, for
fixed time $t$ and for two symmetric orientation of the molecule, namely parallel ($\theta=0$, top panel
corresponding to Eq.~(\ref{axialsym})) and
perpendicular to the field polarization axis, $Z$, ($\theta=\pi/2$, middle and bottom panels corresponding to
Eq.~(\ref{planesymxz}) and Eq.~(\ref{planesymyz}), respectively). $F(t)=0.07$ a.u., $d=2.067$ a.u.\label{potencjal}}
\end{figure}

For a molecule oriented along the field axis the problem possesses axial symmetry. Then switching to the cylindrical coordinates 
($\rho_i,\ \varphi_i,\ z_i$ for $i-$th electron)
one can easily define a symmetry subspace of electron motion. Restricting
the  electrons  to a plane 
(i.e. $\varphi_1-\varphi_2=\pi$) their coordinates in the $C_{2v}$ symmetry subspace are $\rho_1=\rho_2=R$, $z_1=z_2=Z$ and 
the potential energy reduces to
\begin{eqnarray}
\label{axialsym}
V=&-&\frac{2}{\sqrt{R^2+(Z-\frac{d}{2})^2}}\\ \nonumber
&&-\frac{2}{\sqrt{R^2+(Z+\frac{d}{2})^2}}+\frac{1}{2|R|}+2F(t)Z.
\end{eqnarray}
With a molecule oriented perpendicularly with respect to the field there are two $C_{2v}$ symmetry subspaces. 
One subspace is defined in the $xz$ plane the other in the $yz$ plane. In the former case, the electron 
coordinates in the subspace are $(x_1=X,y_1=0,z_1=Z)$ and $(x_2=-X,y_2=0,z_2=Z)$ and the potential energy
 reads 
\begin{eqnarray}
\label{planesymxz}
V=&-&\frac{2}{\sqrt{(X-\frac{d}{2})^2+Z^2}}\\ \nonumber
&&-\frac{2}{\sqrt{(X+\frac{d}{2})^2+Z^2}}+\frac{1}{2|X|}+2F(t)Z.
\end{eqnarray}
And for saddles which are in the $yz$ plane, namely coordinates are $(x_1=0,y_1=Y,z_1=Z)$ and
$(x_2=0,y_2=-Y,z_2=Z)$ and the potential energy is 
\begin{eqnarray}
\label{planesymyz}
V=&-&\frac{4}{\sqrt{\frac{d^2}{4}+Y^2+Z^2}}+\frac{1}{2|Y|}+2F(t)Z.
\end{eqnarray}
 The potential energies Eqs.~(\ref{axialsym}),~(\ref{planesymxz}) and~(\ref{planesymyz}) are
 shown in Fig.~\ref{potencjal} for a set of parameters corresponding to the nitrogen molecule
 (internuclear distance is $d=2.067$ a.u.) and for the field $F(t)=0.07$ a.u. 
 (an intensity of $1.7\times 10^{14}$ W/cm$^2$). The saddles are clearly visible. 

 In a non-sequential double ionization the electrons have to pass sufficiently close to 
 the saddles (exemplified in Fig.~\ref{potencjal} for  molecular orientation being $\theta=0$ and $\theta=\pi/2$). 
For a general orientation
 of a molecule there are two possible channels for non-sequential ionization --- the electrons escape by passing saddles
 situated in a plane defined by the field and molecular axes or in the perpendicular plane. These two channels are equivalent
 when a molecule is parallel to the field and the axial symmetry is restored. 

\begin{figure}[ht]
\includegraphics[height=0.4\textwidth,angle=270]{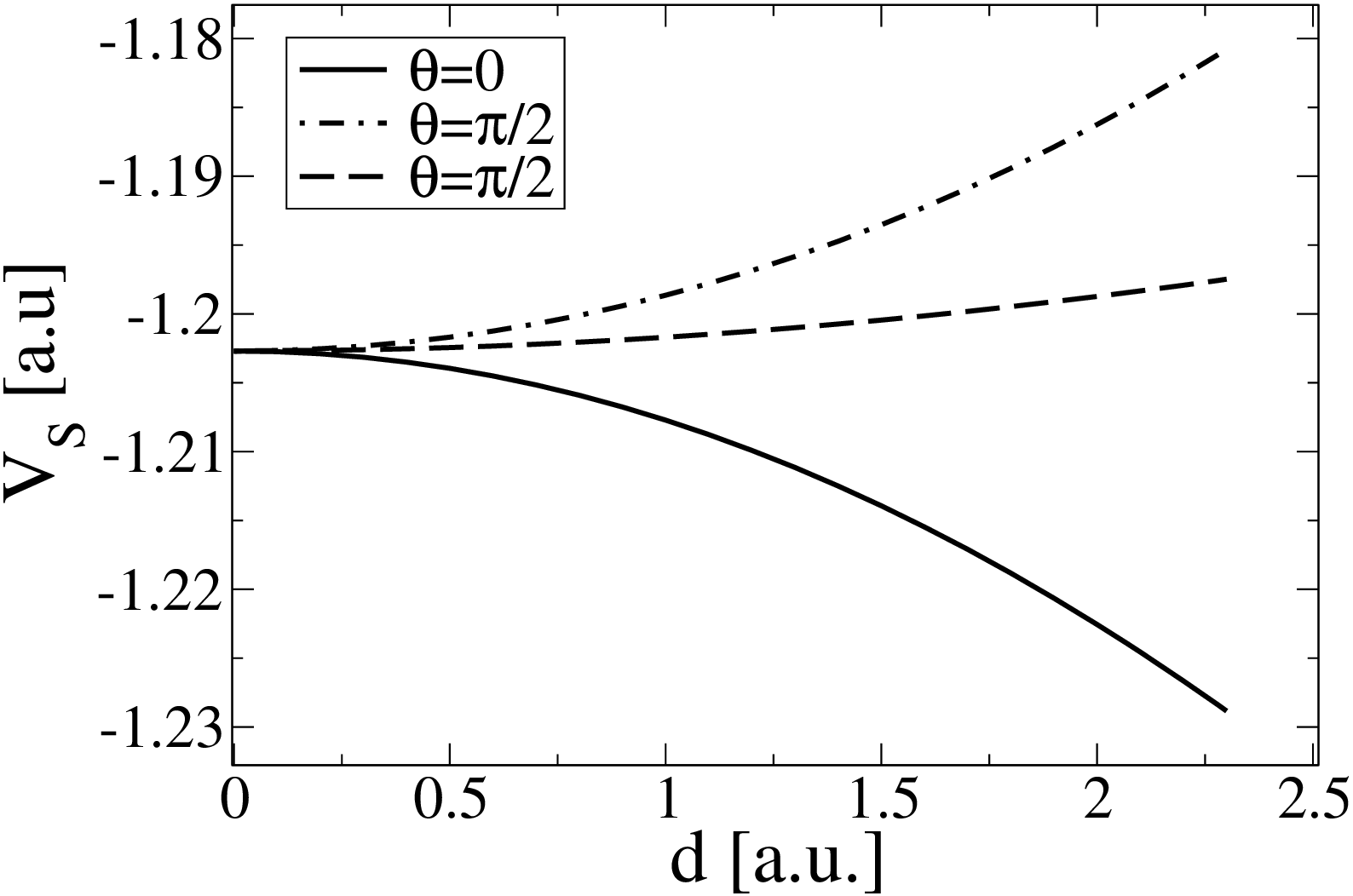}
\includegraphics[height=0.4\textwidth,angle=270]{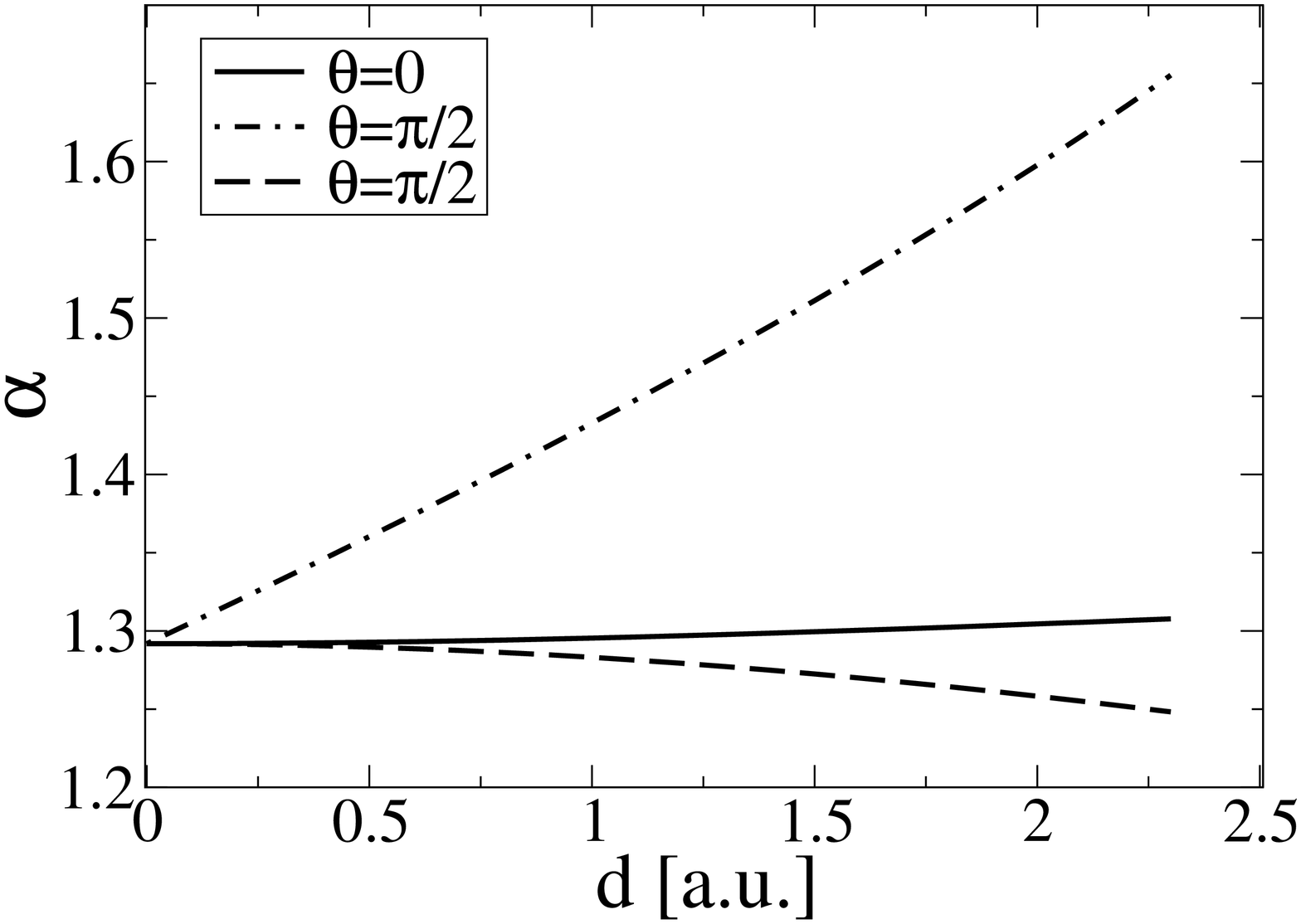}
\caption{Energy of the saddle (top panel) and cross section exponent (bottom panel) as a function of an
increasing internuclear distance, $d$, for two different orientation angles, $\theta$, i.e. $\theta=0$
and $\theta=\pi/2$ (the dashed line correspods to the saddle in $xz$ plane and the dash-dotted line
correspods to the saddle in $yz$ plane, respectively). $F(t)=0.07$
a.u.\label{distance}}
\end{figure}

For a fixed external field local analysis of the saddles reveals a few stable and unstable directions. The latter 
are responsible for either simultaneous double electron escape or single ionization (in such a process one electron leaves 
the molecule while the other returns to the core). Each of unstable directions can be characterized by a Lyapunov
exponent. Knowledge of the Lyapunov exponents allows one, similarly like in the problem of double ionization
without the external field considered many years ago by Wannier~\cite{wannier53,peterkop71,rau84,rost98}, 
to derive the dependence of the cross section on energy close to the threshold, namely,
\begin{equation}
\sigma (E) \propto (E-V_S)^{\alpha},
\end{equation}
where $V_S$ is a saddle energy and the exponent 
\begin{equation}
\alpha=\frac{\sum_i \lambda_i}{\lambda_r},
\end{equation}
where $\lambda_r$ is the Lyapunov exponent of the unstable direction corresponding to the non-sequential 
double ionization and $\lambda_i$ are all other Lyapunov exponents of a saddle~\cite{rost01phe,eckhardt01epl}.

We examine properties of the saddles as the internuclear distance, $d$, increases --- the results are shown in
Fig.~\ref{distance}. 
Starting with $d=0$ (in that case $\theta$ is meaningless since a molecule reduces to an atom)
and increasing $d$ the energy $V_S$ of the saddle corresponding to the molecule 
orientation $\theta=0$ is always the lowest and it decreases. For the other extremal orientation, i.e.
$\theta=\pi/2$, the energies of the saddles (there are two saddles because of the axial symmetry breaking) increase 
and their values are the highest. Analyzing the dependence of $\alpha$ on the internuclear distance 
we see that increasing $d$ the cross section exponent of one of the saddles (corresponding to $\theta=\pi/2$) 
goes up while the other goes down. The exponent for $\theta=0$ slightly increases. 

\begin{figure}[ht]
\includegraphics[height=0.4\textwidth,angle=270]{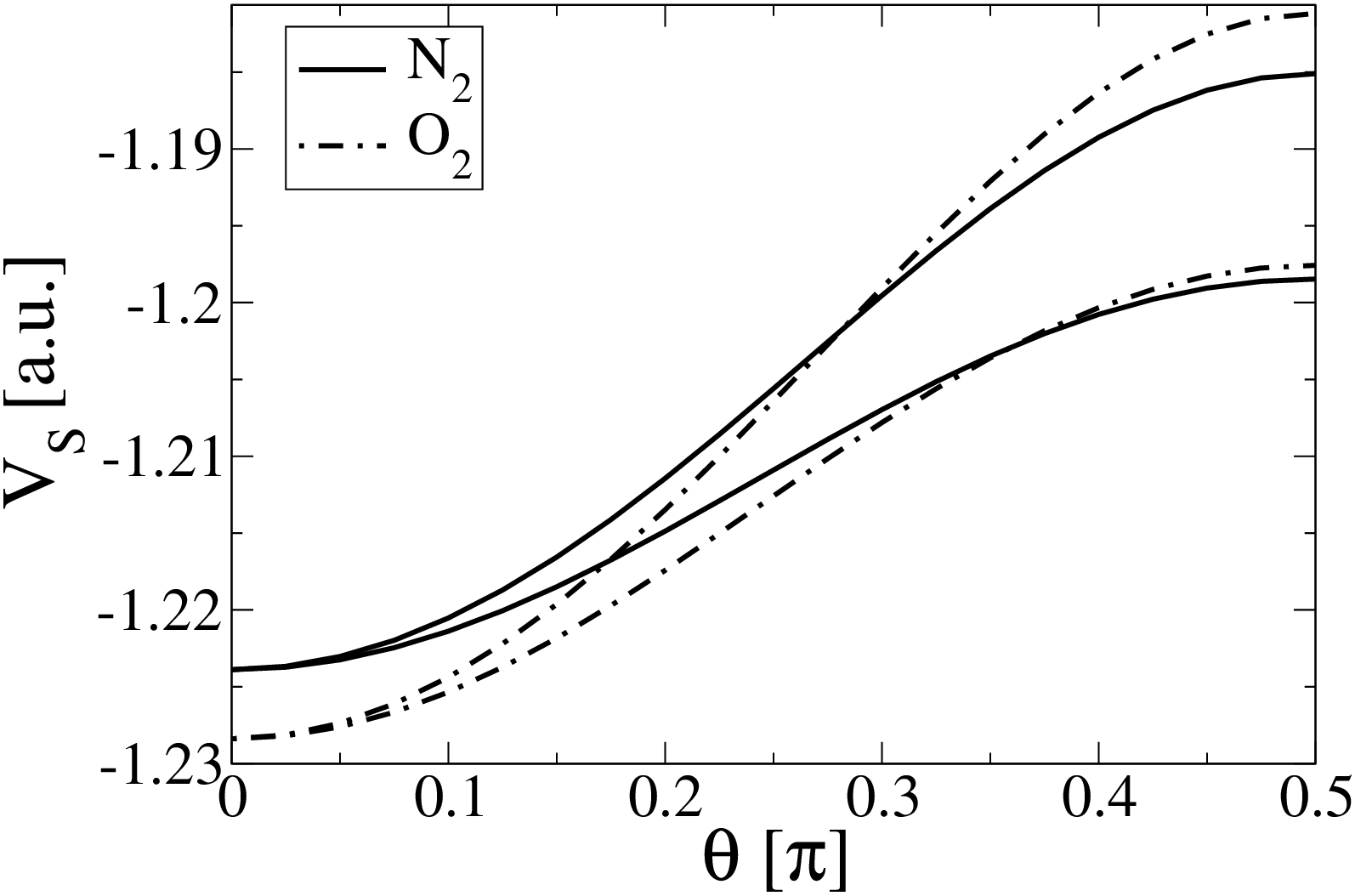}
\includegraphics[height=0.4\textwidth,angle=270]{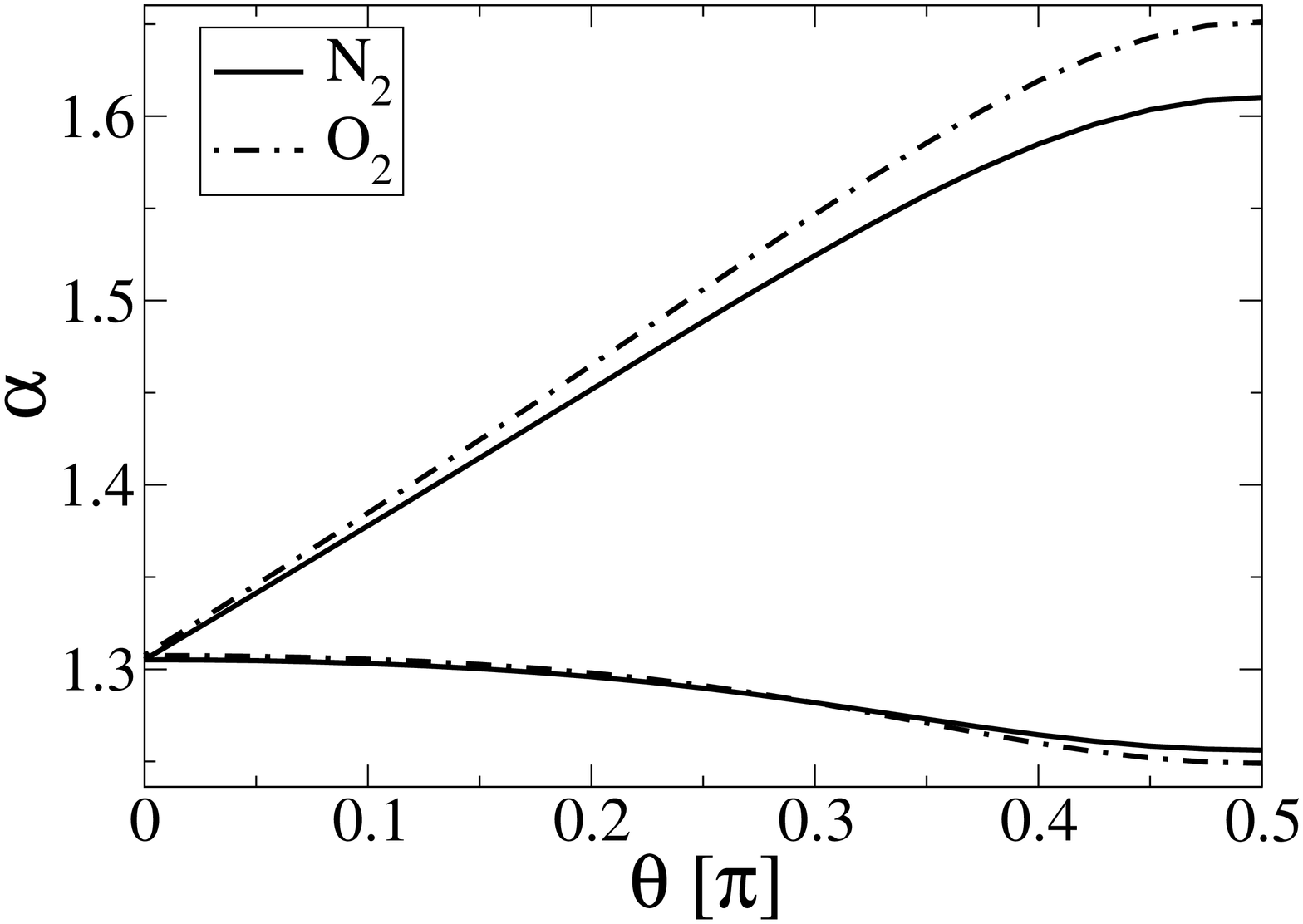}
\caption{Energy of the saddle (top panel) and cross section exponent (bottom panel) as a function of an
orientation angle, $\theta$, for $N_2$ ($d=2.067$ a.u.) and $0_2$ ($d=2.28$ a.u.) molecules.
$F(t)=0.07$ a.u.\label{angle}. (For both molecules the top line corresponds to the saddle in the $yz$
plane, whereas the bottom line to the saddle in the $xz$ plane)}
\end{figure}

Now let us examine how the parameters of the saddles change with the orientation $\theta$ in the case of
$\rm N_2$ ($d=2.067$ a.u.) and $\rm O_2$ ($d=2.28$ a.u.) molecules, see Fig.~\ref{angle}.
The energy $V_S$ of all saddles increases and the exponent $\alpha$ of one member of the saddle pairs 
increases while the other decreases. Fig.~\ref{angle} shows that the parameters for $\rm N_2$ and
$\rm O_2$ molecules are quite similar --- the largest differences between the species are of the order
of few percent. Taking into account that the experimental results \cite{eremina04,miyazaki04} are the statistical 
mixtures of different molecule orientations we may conclude that from the point of view of our local analysis 
of the non-sequential decay channels there should be no differences between $\rm N_2$ and $\rm O_2$.

\section{Numerical Simulations}
\label{sym}

So far we have discussed the local analysis of the potential within the adiabatic assumption. Now we are ready to go 
one step further and perform numerical simulations. In the Coulomb potentials associated with interaction between 
the electrons and nuclei, Eq.~(\ref{pot1}), we introduce a smoothing factor $e$
\cite{su90,reed91} 
to avoid divergence in numerical integration of equation of motions. Then the potential terms read
\begin{eqnarray}
\label{pot3}
V_i=-\frac{1}{\sqrt{(x_i+\frac{d}{2}\sin \theta)^2+y_i^2+(z_i+\frac{d}{2}\cos \theta)^2 +
e}}\cr
-\frac{1}{\sqrt{(x_i-\frac{d}{2}\sin \theta)^2+y_i^2+(z_i-\frac{d}{2}\cos \theta)^2 +
e}}.
\end{eqnarray}
We choose $e=0.01$ which introduces negligible changes of the $V_S$ and $\alpha$ parameters 
presented in Figs.~\ref{distance} and \ref{angle}. 

Assuming  an initial re-scattering event took place, 
two excited electrons
pass close to the nuclei where they interact strongly with each other and with the nuclei. Therefore, it is reasonable to assume,
as mentioned in Sec.~\ref{local} that all the memory of the earlier 
motion is lost. Then the initial state
of the final stage for double ionization is a statistical distribution for two electrons close to the nuclei. Hence, we choose
initial values of positions and momenta with respect to the microcanonical distribution for a given initial energy 
$E$ (it
should be in the range between $-I$ and $-I+3.17U_p$, where $I$ is the ionization energy). The positions are chosen 
microcanonically but with additional conditions, namely $z_i=0$ and $\sqrt{x_i^2+y_i^2}<85$
a.u.~\cite{eckhardt01pra1}. 
We start all simulations at the peak of the laser pulse with the phase of the field chosen randomly. In all simulations an ensamble of $10^5$ trajectories
is used.  

\begin{figure}[t]
\includegraphics[height=0.5\textwidth,width=0.5\textwidth]{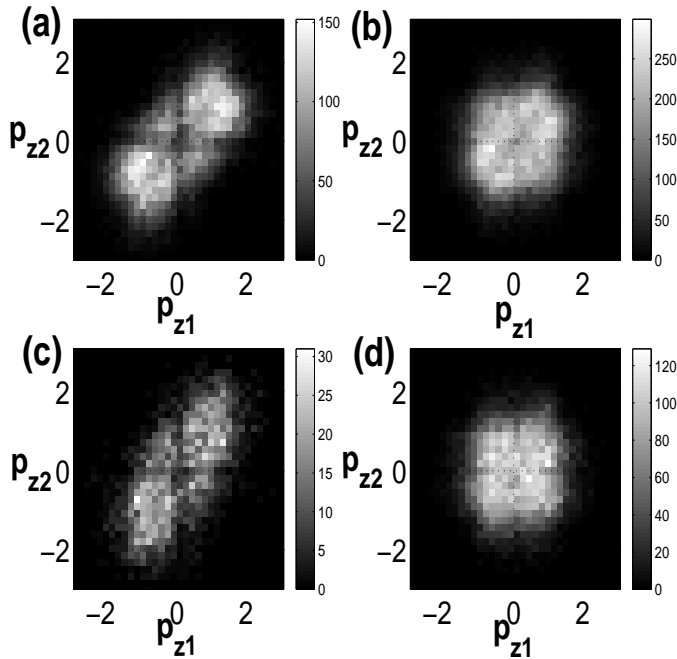}
\caption{Final distribution of the parallel momenta for different initial energy, $E$ and different pulse length
(i.e. number of cycles in the pulse, $n$) in the double ionization of the nitrogen molecule parallel to the
field axis\label{theta_zero}. $F(t)=0.07$ a.u., $\omega=0.057$ a.u., $d=2.067$ a.u. and $\theta=0$. (a) $E=-0.3$ a.u.,
$n=2$; (b) $E=-0.3$ a.u., $n=26$; (c) $E=-0.6$ a.u., $n=2$ and (d) $E=-0.6$ a.u., $n=26$.}
\end{figure}

\begin{figure}[ht]
\includegraphics[height=0.4\textwidth,angle=270]{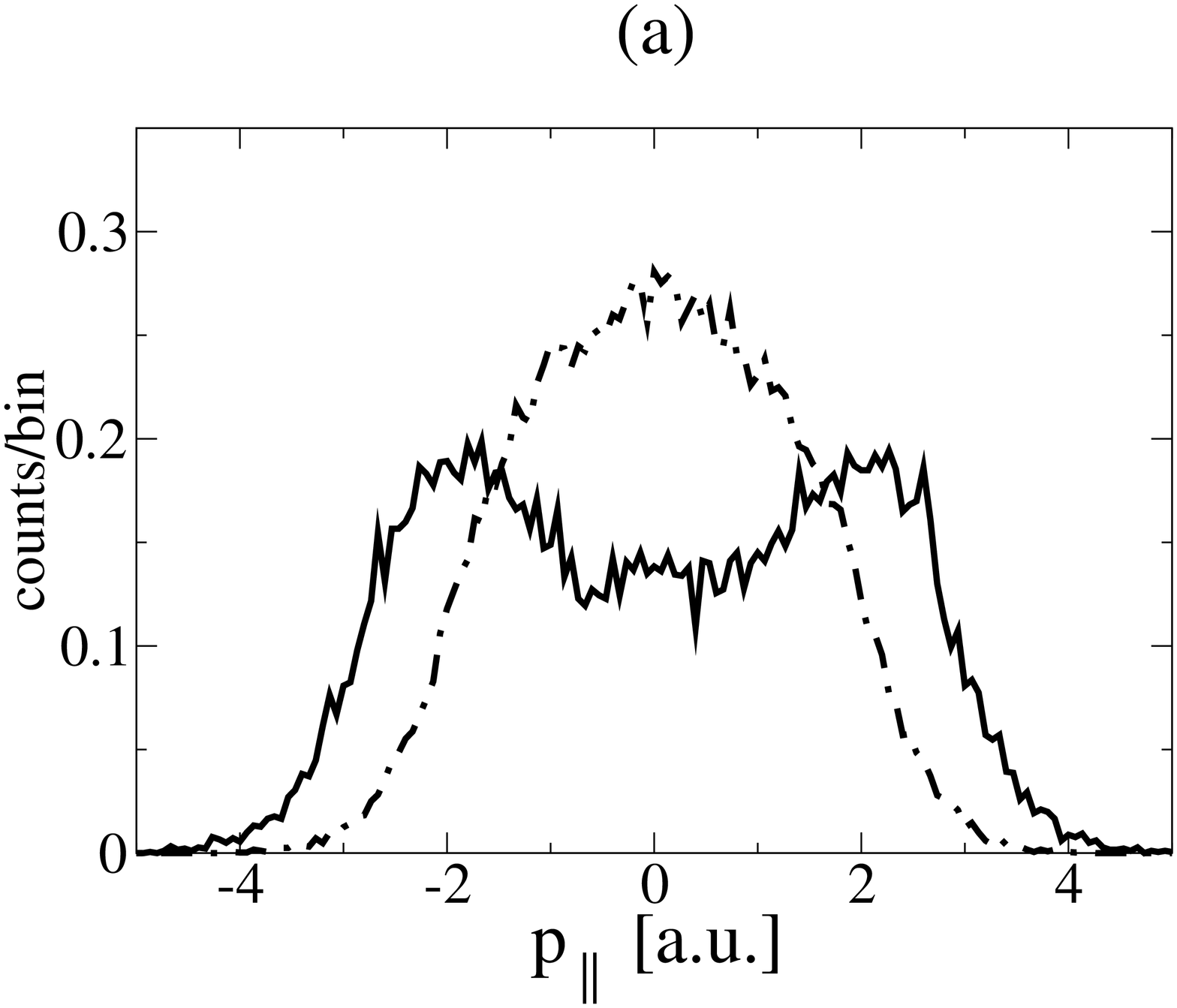}
\includegraphics[height=0.4\textwidth,angle=270]{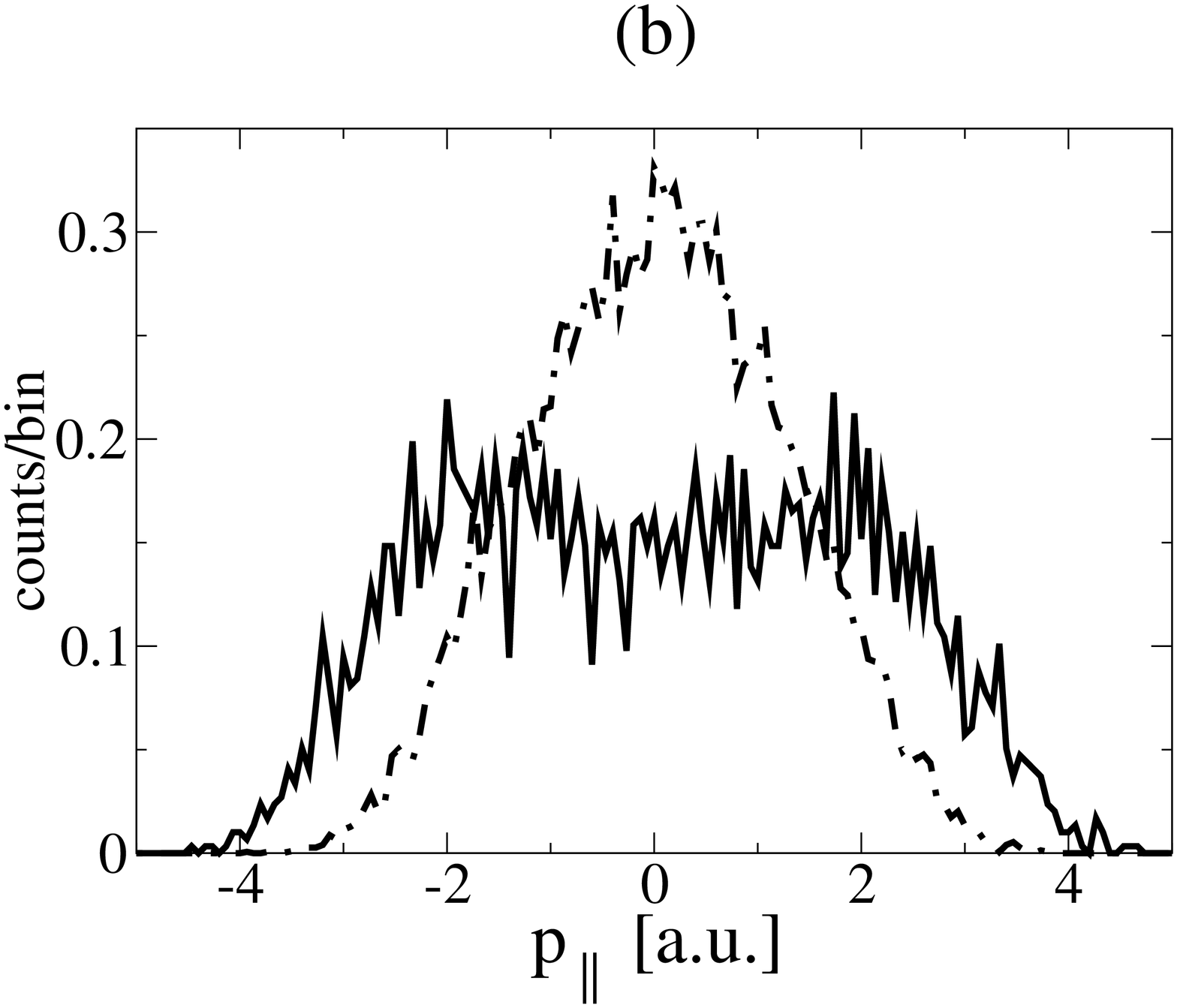}
\caption{Final distribution of ion parallel momenta corresponding to Fig.~\ref{theta_zero}.
 $F(t)=0.07$ a.u., $\omega=0.057$ a.u., $d=2.067$ a.u. and $\theta=0$. (a) $E=-0.3$ a.u., $n=2$
 - solid line and $n=26$ - broken line; (b) $E=-0.6$ a.u., $n=2$ - solid line and $n=26$ - broken line.}
\label{jony}
\end{figure}

Distributions of the final electron momenta parallel to the field axis for different initial 
energy $E$ and for different pulse durations are presented in Fig.~\ref{theta_zero} . All data in the figure correspond 
to the $\rm N_2$ molecule oriented along the field axis. For very short laser pulses ($n=2$ cycles) signatures 
of simultaneous electron escape are clearly visible --- the distributions are localized along the diagonals indicating 
that the electrons escape predominately by passing close to the saddles analyzed in Sec.~\ref{local}. For longer pulses
the distributions change their character. The first and fourth quadrants of the panels 
become strongly populated implying that 
a number of sequential decays significantly increases. The reason for that is quite obvious.
After re-scattering when a highly excited two electron complex is created, there are two dominant scenarios within the first half cycle of the field: 
a non-sequential double escape or a single ionization. During the next cycles, unless the molecule is
already doubly ionized, 
we are left with a singly ionized molecule
which may survive to the end of the pulse or the second electron escapes and that corresponds to a sequential double ionization.
The longer pulse duration the more sequential ionizations which may easily overcome the number of 
non-sequential events. Thus for sufficiently long pulses, even though the re-scattering scenario is involved in the double 
ionization process, the momentum distributions will show signatures of the sequential electron escape. 
If the initial energy $E$ is much higher than the minimal energy of a saddle 
($V_S\approx -1.2$~a.u.) then probability of non-sequential ionization is bigger and even if the pulse duration is quite 
long (e.g. $n=26$ in panel (b) of Fig.~\ref{theta_zero}) the signatures of non-sequential process remain (contrary to panel 
(d) where for $E=-0.6$~a.u. such signatures are not visible). 
In Fig.~\ref{jony} one can find the distributions of ion parallel momenta that correspond to the data presented in 
Fig.~\ref{theta_zero}. The distributions, as expected, are much narrower in the case when sequential ionization dominates 
than in the case when non-sequential process does. 

\begin{figure}[t]
\includegraphics[height=0.3\textwidth,width=0.3\textwidth]{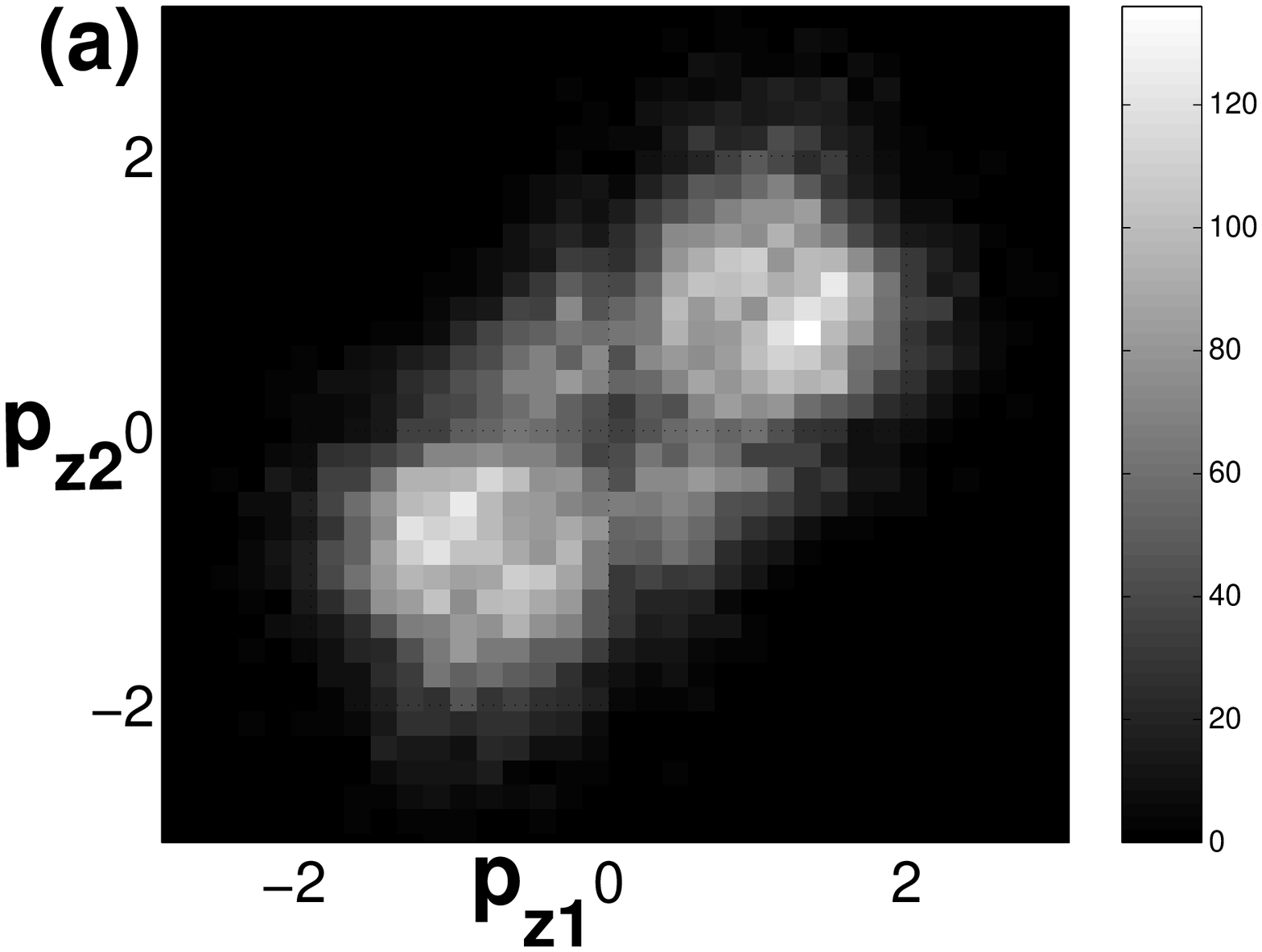}
\includegraphics[height=0.3\textwidth,width=0.3\textwidth]{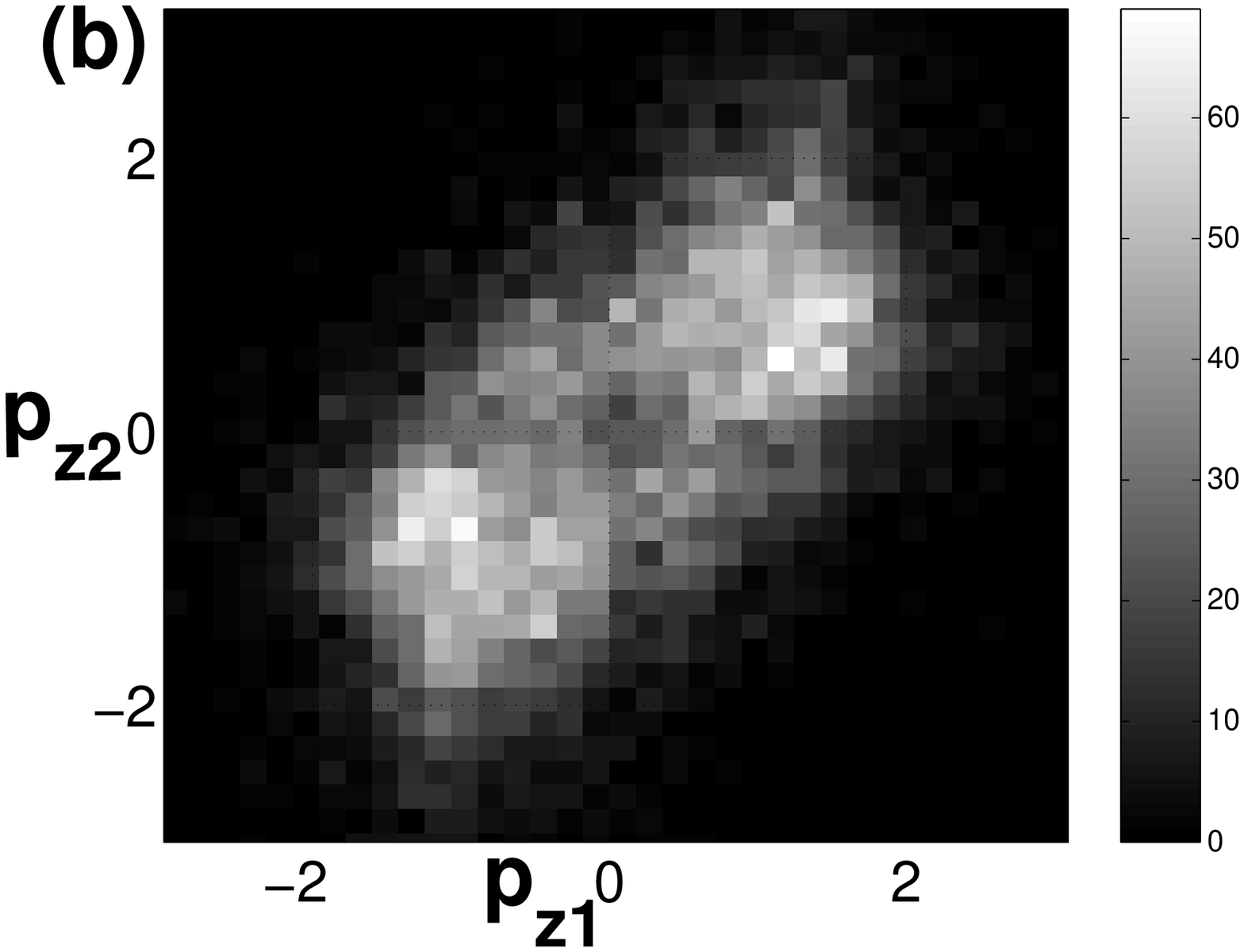}
\caption{Final distribution of the parallel momenta for the nitrogen molecule at non-zero angle
to the field axis.\\ $F(t)=0.07$ a.u., $\omega=0.057$ a.u., $d=2.067$ a.u., $E=-0.3$ a.u. and
$n=2$. (a) $\theta=\pi/4$ and (b) $\theta=\pi/2$}
\label{dangle}
\end{figure}

When we change the orientation of the molecule the distributions do not change significantly as one can see comparing 
Fig.~\ref{dangle} and Fig.~\ref{theta_zero}. For the field amplitude used in the simulations (which corresponds also
to the experiment \cite{eremina04}) the saddles for non-sequential process are lying quite away from the nuclei and the 
positions and other parameters of the saddles change slightly only with a change of $\theta$ --- compare Fig.~\ref{angle}. 

\begin{figure}[ht]
\includegraphics[height=0.3\textwidth,width=0.3\textwidth]{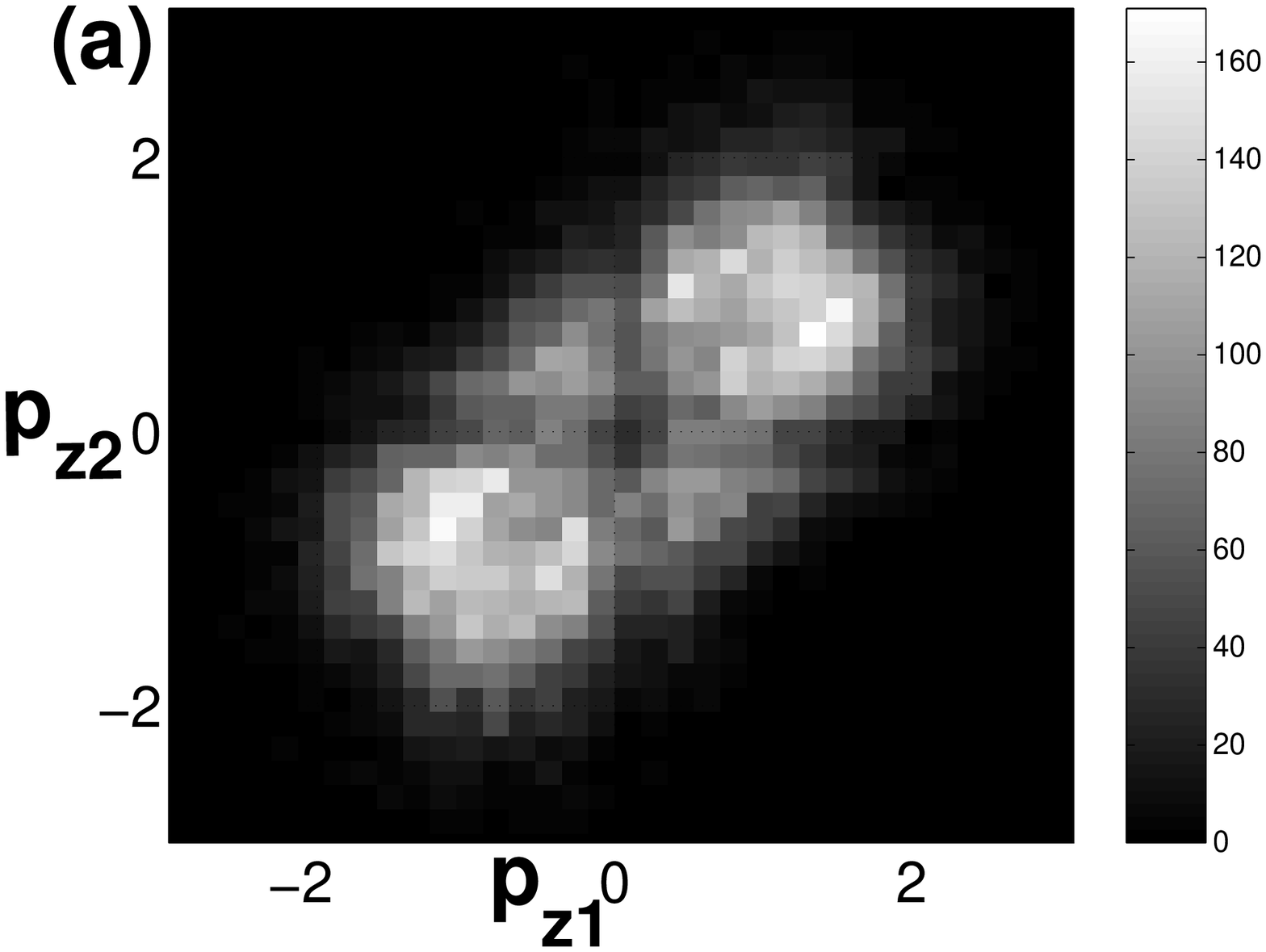}
\includegraphics[height=0.3\textwidth,width=0.3\textwidth]{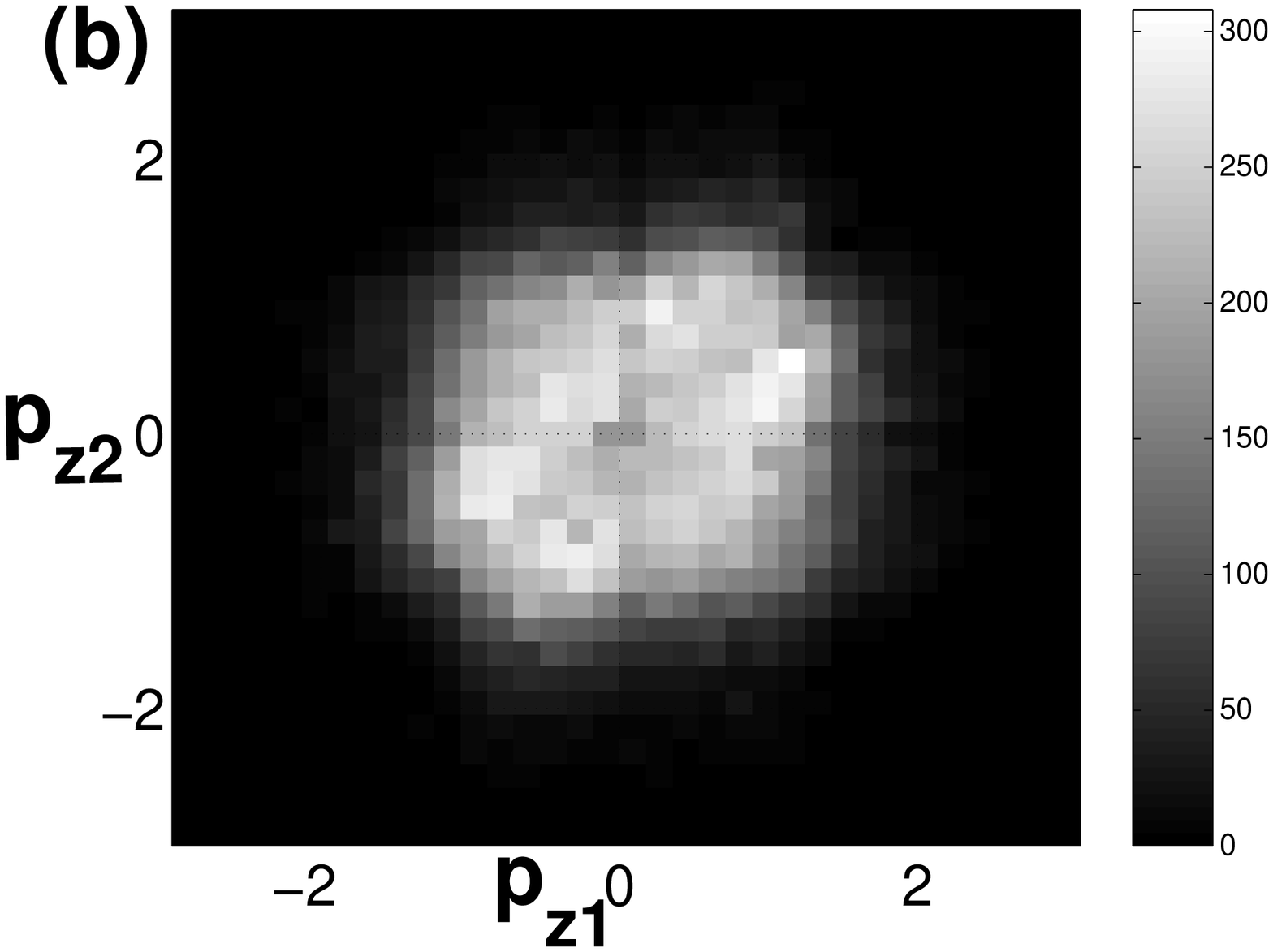}
\caption{Final distribution of the parallel momenta for the oxygen molecule parallel
to the field axis for different pulse length. $F(t)=0.07$ a.u., $\omega=0.057$ a.u., $d=2.28$ a.u., $E=-0.3$ a.u.
and $\theta=0$. (a) $n=2$ and (b) $n=26$.}
\label{oxygen_dist}
\end{figure}

Finally in Fig.~\ref{oxygen_dist} we show data for the $\rm O_2$ molecule. As expected, from the local analysis presented
in Sec.~\ref{local}, for the same initial energies $E$ and pulse durations as in the case of the $\rm N_2$ molecule observed
momentum distributions are very similar, compare Fig.~\ref{theta_zero}.
In Eremina {\it et al., } experiment \cite{eremina04} the results for $\rm O_2$ differ from the data for $\rm N_2$. 
The latter shows signatures of simultaneous electron escape while for $\rm N_2$
it seems that sequential process dominates. 
Our analysis indicates that there is practically no difference between $\rm N_2$ and $\rm O_2$
if similar initial conditions for highly excited complex are chosen.
It strongly suggests that the differences between the observations in the two
experiments are due to differences in the early stages of the excitation
process and the nature of the compound state before the final decay towards
multiple ionization.

Our analysis also indicates that
 to increase the ratio of non-sequential to sequential ionizations
very short pulses should be used. It opens a possibility for observation of clear signatures of simultaneous electron escape 
for $\rm O_2$ molecule, too.

\section{Summary}
\label{summary}

To summarize, we have performed a purely classical analysis of the final stage of the non-sequential 
double ionization of molecules in the strong laser field. It is based on the fact that all 
trajectories leading to the non-sequential escape of electrons have to pass close to the saddles 
in the potential that is formed when the laser pulse arrives 
\cite{eckhardt01pra1,eckhardt01pra2,eckhardt01epl,eckhardt03jpb}. 

We have started with the local analysis of the potential within the adiabatic approximation, and we have
identified and described the saddles. Thereafter, we have shown results of  numerical simulations.
The later allow us to draw two conclusions: i)~From the point of view of classical analysis 
there is no difference between nitrogen and oxygen molecules in a sense that both of them can show signatures
of simultaneous double escape. ii)~Orientation of the molecule with respect to the field axis 
does not influence significantly the final momentum distribution for the initial energy range considered in
the paper. 
 
 The numerical results and their interpretation suggests strongly
 that shorter laser pulses
 should lead to an increase of the ratio of non-sequential to sequential ionizations. That suggestion could be tested experimentally.
 Higher relative efficiency of non-sequential process would be visible in the momentum
distributions as a more pronounced symmetrical escape of the electrons.

\section{Acknowledgments}
We are grateful to E. Eremina for making the experimental results accessible before publication.

This work  was partly supported by the Polish Ministry of 
 Scientific Research Grant 
 PBZ-MIN-008/P03/2003 and by the Deutsche Forschungsgemeinschaft.

\begin{thebibliography}{10}

\bibitem{silap93}
{\em Super-Intense Laser-Atom Physics}, {\em Proceedings of the NATO
  Advanced Research Workshop, Han-sur-Lesse, Belgium, 1993}, edited by B.
  Piraux, A. L'Huillier, and K. Rz\c{a}\.zewski (Plenum Press, New York, 1993).

\bibitem{silap00}
{\em Super-Intense Laser-Atom Physics}, {\em Proceedings of the NATO
  Advanced Research Workshop, Han-sur-Lesse, Belgium, 2000}, edited by B.
  Piraux and K. Rz\c{a}\.zewski (Kluwer Academic Publishers, Dordrecht, 2001).

\bibitem{schafer93}
K.~J. Schafer, B. Yang, L.~F. DiMauro, and K.~C. Kulander, Phys. Rev. Lett.
  {\bf 70},  1599  (1993).

\bibitem{yang93}
B. Yang, K. J. Schafer, B. Walker, K. C. Kulander, P. Agostini, 
L. F. DiMauro, Phys. Rev. Lett. {\bf 71},  3770  (1993).

\bibitem{eckhardt01pra1}
K. Sacha and B. Eckhardt, Phys. Rev. A {\bf 63},  043414  (2001).

\bibitem{eckhardt01pra2}
K. Sacha and B. Eckhardt, Phys. Rev. A {\bf 64},  053401  (2001).

\bibitem{eckhardt01epl}
B. Eckhardt and K. Sacha, Europhys. Lett. {\bf 56},  651  (2001).

\bibitem{eckhardt03jpb}
K. Sacha and B. Eckhardt, J. Phys. B: At. Mol. Opt. Phys. {\bf 36},  3923
  (2003).

\bibitem{corkum93}
P.~B. Corkum, Phys. Rev. Lett. {\bf 71},  1994  (1993).

\bibitem{kulander93}
K.~C. Kulander, K.~J. Schafer, and J.~L. Krause,  in {\em Super-Intense
  Laser-Atom Physics}, {\em Proceedings of the NATO Advanced Research Workshop,
  Han-sur-Lesse, Belgium, 1993}, edited by B. Piraux, A. L'Huillier, and K.
  Rz\c{a}\.zewski (Plenum Press, New York, 1993).

\bibitem{cornaggia98}
C. Cornaggia and P. Hering, J. Phys. B: At. Mol. Opt. Phys. {\bf 31},  L503
  (1998).

\bibitem{guo98}
C. Guo, M. Li, J.~P. Nibarger, and G.~N. Gibson, Phys. Rev. A {\bf 58},  R4271
  (1998).

\bibitem{guo00}
C. Guo, M. Li, J.~P. Nibarger, and G.~N. Gibson, Phys. Rev. A {\bf 61},  033413
   (2000).

\bibitem{eremina04}
E. Eremina, X. Liu, H. Rottke, W. Sandner, M. G. Sch\"atzel, A. Dreischuh, G. G. Paulus, H. Walther,
 R. Moshammer and J. Ullrich, Phys. Rev. Lett. {\bf 92},  173001  (2004).

\bibitem{miyazaki04}
K. Miyazaki, T. Shimizu and D. Normand, J. Phys. B: At. Mol. Opt. Phys. {\bf 37},  753
  (2004).

\bibitem{wannier53}
G.~H. Wannier, Phys. Rev. {\bf 90},  817  (1953).

\bibitem{peterkop71}
R. Peterkop, J. Phys. B: At. Mol. Phys. {\bf 4},  513  (1971).

\bibitem{rau84}
A.~R.~P. Rau, Phys. Rep. {\bf 110},  369  (1984).

\bibitem{rost98}
J.~M. Rost, Phys. Rep. {\bf 297},  271  (1998).

\bibitem{rost01phe}
J.~M. Rost, Physica E {\bf 9},  467  (2001).

\bibitem{su90}
Q. Su, J.~H. Eberly, and J. Javanainen, Phys. Rev. Lett. {\bf 64},  862
  (1990).

\bibitem{reed91}
V.~C. Reed, P.~L. Knight, and K. Burnett, Phys. Rev. Lett. {\bf 67},  1415
  (1991).

\end{thebibliography}

\end{document}